\documentclass[conference]{IEEEtran}
\IEEEoverridecommandlockouts
\ifCLASSOPTIONcompsoc
  \usepackage[nocompress]{cite}
\else
  \usepackage{cite}
\fi
\usepackage{hyperref}

\ifCLASSINFOpdf
\else
\fi
\usepackage{amsmath}
\ifCLASSOPTIONcompsoc
  \usepackage[caption=false,font=footnotesize,labelfont=sf,textfont=sf]{subfig}
\else
  \usepackage[caption=false,font=footnotesize]{subfig}
\fi

\usepackage{booktabs, tabulary}       
\usepackage{amsfonts}       
\usepackage{nicefrac}       
\usepackage{microtype}      
\usepackage{lipsum}
\usepackage{algorithm}
\usepackage{mathtools}
\usepackage[noend]{algpseudocode}
\usepackage{float}

\begin{document}

\title{Topological Data Analysis for Portfolio Management of Cryptocurrencies \\
\thanks{The research in \autoref{sec:problemStatement}, and \autoref{section:dataset} was supported by the Ministry of Education and Science of the Russian Federation (Grant no. 14.756.31.0001). Other sections were supported by the Mexican National Council for Science and Technology (CONACYT), 2018-000009-01EXTF-00154.}
}

\author{\IEEEauthorblockN{Rodrigo Rivera-Castro, Polina Pilyugina, Evgeny Burnaev}
\IEEEauthorblockA{\textit{Skolkovo Institute of Science and Technology} \\
Moscow, Russia \\
rodrigo.riveracastro@skoltech.ru}

}

\maketitle

\begin{abstract}
  Portfolio management is essential for any investment decision. Yet, traditional methods in the literature are ill-suited for the characteristics and dynamics of cryptocurrencies. This work presents a method to build an investment portfolio consisting of more than 1500 cryptocurrencies covering 6 years of market data. It is centred around Topological Data Analysis (TDA), a recent approach to analyze data sets from the perspective of their topological structure. This publication proposes a system combining persistence landscapes to identify suitable investment opportunities in cryptocurrencies. Using a novel and comprehensive data set of cryptocurrency prices, this research shows that the proposed system enables analysts to outperform a classic method from the literature without requiring any feature engineering or domain knowledge in TDA. This work thus introduces TDA-based portfolio management of cryptocurrencies as a viable tool for the practitioner.
\end{abstract}


\section{Originality and Value}
This research presents a portfolio approach for investing in cryptocurrencies using a method inspired by Topological Data Analysis techniques for time series data. The proposed method has been validated with six years of market data from 1561 cryptocurrencies.

The contributions cover the areas of portfolio management and time series analysis. They are suited for individuals with domain knowledge of time series and cryptocurrencies but limited understanding of machine learning methods. This work innovates by outlining:
\begin{enumerate}
\item A practical case of porfolio management for cryptocurrencies,
\item An application of time series using Topological Data Analysis,
\item A novel data set providing daily market data for 1561 cryptocurrencies and spanning 6 years until early July 2019,
\item For reproducibility purposes, an implementation and data set available for download\footnote{\url{https://github.com/rodrigorivera/icdm_bda19}}.
\end{enumerate}

\section{Problem Statement}\label{sec:problemStatement}
Cryptocurrencies have become a focal point of attention for investors, regulators, media and the general population in the last 10 years, \cite{Corbet_undated-au}. They can be described as a system to settle peer-to-peer payments through the use of electronic cash and an algorithm based on encryption enabling to keep track of all historic transactions in the system.

The existence of a centralized ledger enables both parties to carry out financial transactions without intermediaries such as financial institutions. As such, cryptocurrencies do not have oversight from a higher authority or have a material representation. Similarly, there can be infinitely many cryptocurrencies. Unlike traditional financial investments, they do not have an underlying tangible asset or the backing of a government or a company.
\cite{Corbet2018-td} sees three reasons for the growth in popularity of cryptocurrencies, their low transaction costs, peer-to-peer design and lack of governmental regulation. As a result, they argue that this has led to a surge in trading volume, the price of cryptocurrencies and an increase in volatility.

Although at the moment of this writing there are more than 2000 cryptocurrencies available in the market with a total capitalization of \$279.5 billion, Bitcoin remains the market leader with a capitalization of \$193.4 billion. Its valuation and the market for cryptocurrencies have been under constant change. For example, in the space of one year, from October 2016 to October 2017, the market capitalization of Bitcoin went from \$10.1 billion to \$79.7 billion. As a result, it was possible to achieve a 680\% return on investment per year, a figure that according to \cite{Corbet_undated-au} cannot be offered by any other asset. At its peak in December 2017, Bitcoin reached the price of \$19,500. As of August of 2019, its price has been around \$10,800 and in early 2019, it reached a bottom of \$3,400.

The combination of a high volatility, see \autoref{fig:tda:market_volatility_20190820}, and wide variety of currencies, see \autoref{fig:tda:overview_cryptocurrencies}, makes it difficult to identify investment opportunities. At the same time, frameworks to optimize portfolios are highly sensitive to estimation errors and thus poorly suited to make investment decisions with cryptocurrencies, \cite{Platanakis2019-jk}.

The limited work on portfolio theory for cryptocurrencies and their broader societal potential motivate this research. Its objective is to present a portfolio selection method consisting of a technique for (1) time series processing using Topological Data Analysis and a (2) a framework from the marketing literature for customer profiling used as criteria to establish the asset allocation, \cite{Zhang2015-wa}. This work seeks to achieve superior performance over a 1/N 'buy and hold' strategy for a portfolio of cryptocurrencies.

\begin{figure}[!ht]
  \begin{center}
    \includegraphics[width=\columnwidth]{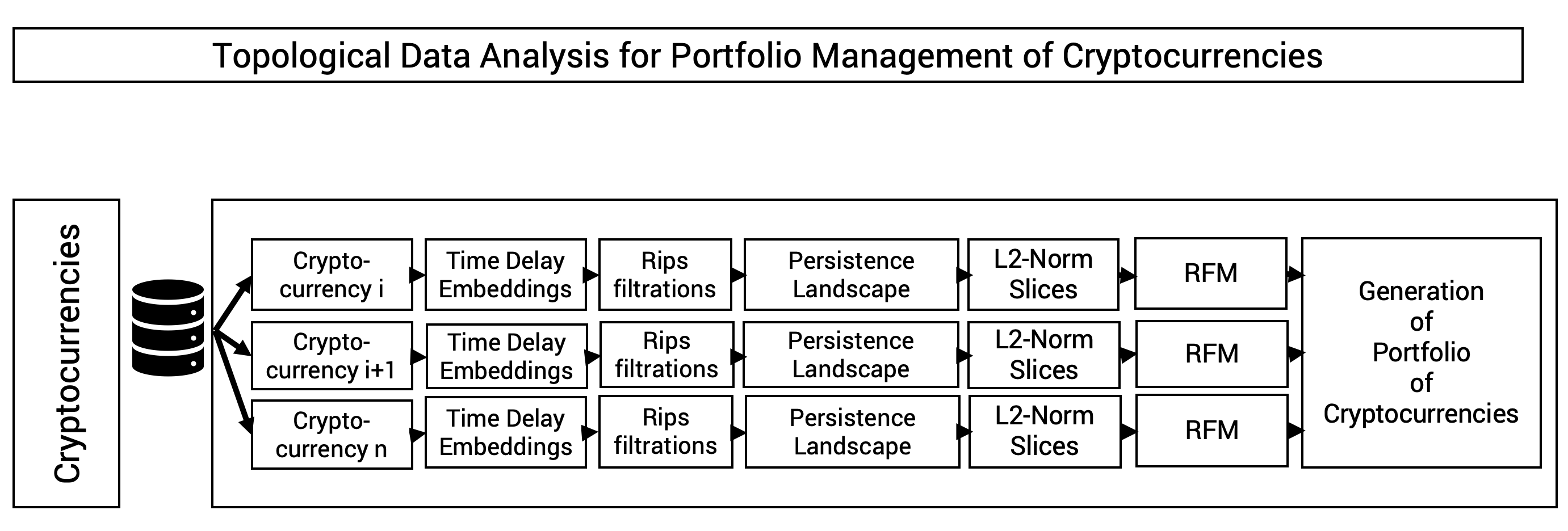}
  \end{center}
  \caption{Overview of the proposed pipeline for portfolio management}
  \label{fig:overview_portfolio}
\end{figure}

\section{Related Work}\label{section:literaturereview}
The field of cryptocurrencies is relatively new. However, the work of \cite{Gidea2018-ah} is important to highlight. They also propose a pipeline using TDA for cryptocurrencies to identify major changes in the price of a currency. Their assessment was made with 3 cryptocurrencies and for a limited period of two years. 
More broadly, authors from other communities have tried to assess the market of cryptocurrencies. For example, \cite{Platanakis2019-jk} proposed a portfolio method based on estimating risk on a selection of cryptocurrencies consisting of Bitcoin, Litecoin, Ripple and Dash. Others such as \cite{Corbet_undated-au} tried to develop a system to understand the market of cryptocurrencies as an asset class with a basket of the most popular ones such as Bitcoin, Ethereum, Litecoin, Dash and Ripple. Similarly, \cite{Corbet2018-td} tried to establish a relation between cryptocurrencies an other popular asset classes.

Other authors such as \cite{Akcora_undated-ew} have tried to use predictive methods to estimate the price of Bitcoin and applied Topological Data Analysis techniques for ransomware detection in the blockchain, \cite{Akcora2019-rw}. Other works have sought to develop analytics for the Blockchain such as \cite{Akcora2018-gl}. With the exception of \cite{Gidea2018-ah}, this work is not familiar with further literature developoing methods to evaluate cryptocurrencies using Machine Learning methods or TDA.

Broadly, Topological Data Analysis (TDA) can be seen as a combination of various statistical, computational, and topological methods with the objective of finding shape-like structures in the data. For a detailed introduction, the reader is invited to consult \cite{books/daglib/0025666}, \cite{boissonnat:hal-01615863} or \cite{chazal2017introduction}.

Similarly, this work draws inspiration from the Recency Frequency Monetary framework (\textit{RFM}) popular in the marketing literature. Its objective is to to provide answers to the questions 'When was the last time that an individual was active?', 'How frequently has an individual been active for a given time period?', 'How much revenue has she generated?'. 
For each of these questions, a score is calculated. As a result, a three-digit number is obtained. The score is indicative for the type of individual; low scores highlight valuable users, \cite{doi:10.1177/1094670506293810}. For an overview, the works of \cite{Zaki2016-oj}, \cite{Wubben2008-yf} and \cite{Rivera-Castro2019-qk} can be consulted. 

Throughout the literature, a noticeable constant is the limited scope of the studies due to a lack of sufficient data. An objective of this study is to assess the largest possible subset of cryptocurrencies, 1561, over a long period of time, six years of data. 
Similarly, this work differentiates itself by using a $L$-norm and seeking to extract features to estimate a suitable portfolio allocation. 

\section{Data Set}\label{section:dataset}
\begin{figure}[!ht]
  \begin{center}
    \includegraphics[width=\columnwidth]{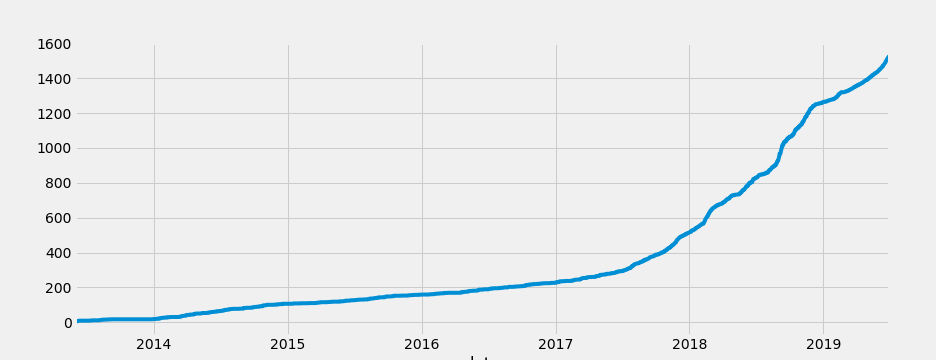}
  \end{center}
  \caption{Overview of number of cryptocurrencies by day}
  \label{fig:tda:overview_cryptocurrencies}
\end{figure}

The data set consists of six years of daily market data representing the USD value of 1561 cryptocurrencies. Given the recency and popularity of this market, the number of cryptocurrencies has been constantly increasing over time, this can be seen for example in \autoref{fig:tda:overview_cryptocurrencies}. 
Although experiments were done on all cryptocurrencies from the provided data set, this research will center on discussing Bitcoin due to its market relevancy and history.

\begin{figure}[!ht]
  \begin{center}
    \includegraphics[width=\columnwidth]{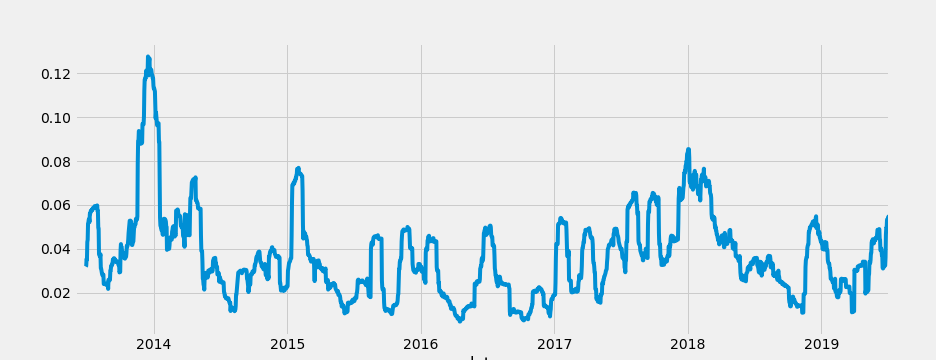}
  \end{center}
  \caption{Volatility of Bitcoin for the period 2013-05-01 to 2019-06-30}
  \label{fig:tda:btc_volatility_20190501_20190701}
\end{figure}

\begin{figure}[!ht]
  \begin{center}
    \includegraphics[width=\columnwidth]{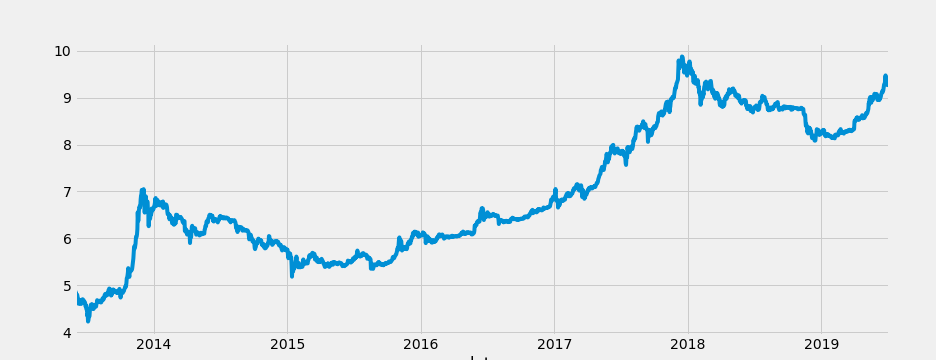}
  \end{center}
  \caption{Log USD values of Bitcoin for the period 2013-05-01 to 2019-06-30}
  \label{fig:tda:btc_log_close_20190501_20190630}
\end{figure}

\begin{figure}[!ht]
  \begin{center}
    \includegraphics[width=\columnwidth]{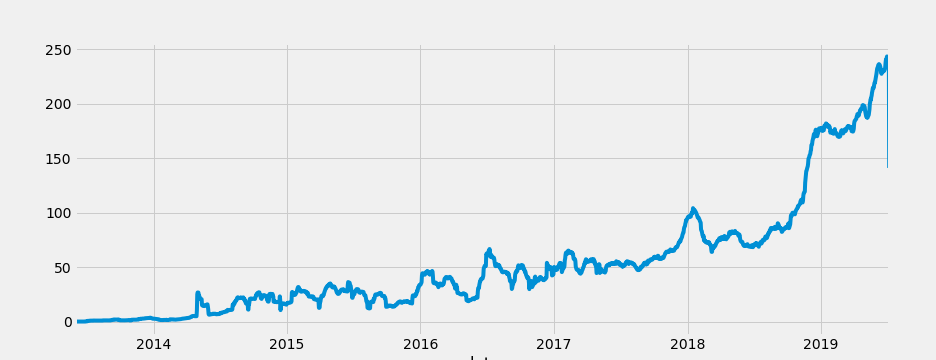}
  \end{center}
  \caption{Aggregated volatility of 1561 cryptocurrencies for the period 2013-05-01 to 2019-06-30}
  \label{fig:tda:market_volatility_20190820}
\end{figure}

The historic volatility of Bicoin since 2013 can be seen in \autoref{fig:tda:btc_volatility_20190501_20190701} and compared against the market volatility in \autoref{fig:tda:market_volatility_20190820}. It is visible that Bitcoin is significantly more stable than its peers. 
Yet, Bitcoin remains a highly profitable asset class, as seen on the historic log USD value of the cryptocurrency in \autoref{fig:tda:btc_log_close_20190501_20190630}.

\begin{figure}[!ht]
  \begin{center}
    \includegraphics[width=\columnwidth]{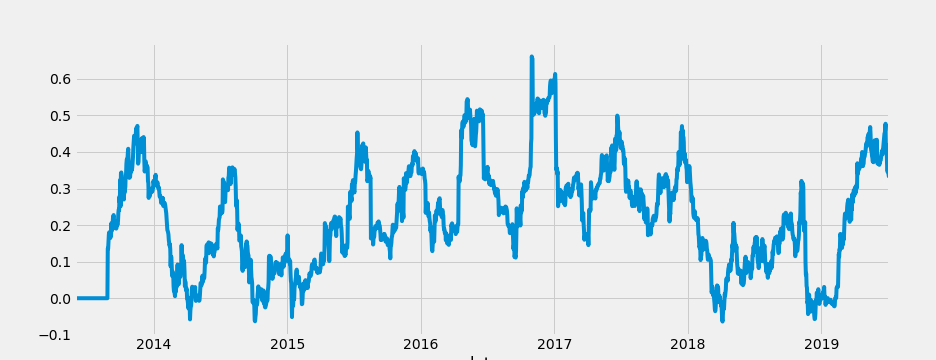}
  \end{center}
  \caption{60 days rolling Sharpe Ratio for Bitcoin for the periods 2013-05-01 to 2019-06-30}
  \label{fig:tda:btc_sharpe_ratio_20190501_20190701}
\end{figure}

This is further emphasized in \autoref{fig:tda:btc_sharpe_ratio_20190501_20190701} depicting the historic rolling 60 days Sharpe Ratio of Bitcoin. This is a measure of the risk-adjusted return of a financial portfolio with high values being considered as desirable. One can see that the Sharpe Ratio of Bitcoin has been mostly positive through the six years of data available.

\section{TDA for Portfolio Management of Cryptocurrencies}\label{sec:tda_user_profiling}
\autoref{fig:overview_portfolio} provides a global overview of the architecture proposed in this work. This section details the elements integrating this system. The proposed pipeline is applied to each cryptocurrency as following:

\begin{align*}
\text{Time Series} \rightarrow \text{Time Delay Embeddings} \rightarrow\\
\text{Rips Complex} \rightarrow \text{Homology} \rightarrow \\
\text{Landscapes} \rightarrow \text{L2-Norms} \rightarrow \\
\text{RFM} \rightarrow \text{Score}
\end{align*}
 or more formally
\begin{align*}
f_i \rightarrow Z_j \rightarrow \\
\{R(Z_j, \epsilon)\}_\epsilon \rightarrow {H_*(R(Z_j,\epsilon))}_\epsilon  \rightarrow \\
\{\lambda^j_k\}_{k \geq 0} \rightarrow \{ \|\lambda^j_k\|_2\}_{k \geq 0} \rightarrow \\
\text{RFM} \rightarrow \mathbb{N}^3 
\end{align*}.

\subsection{Time Series}\label{subsec:pipeline:time_series}
A time series can be defined as a sequence of tuples $(x_0,f_0),\ldots,(x_k,f_k)$, where $x \in \mathbb{N}$ such that at index $i$, a function $f_i = f(x_i)$; $f_i$ represents one cryptocurrency; for example, Bitcoin.

\subsection{Time Delay Embedding}\label{subsec:pipeline:tde}
The idea behind TDA is to build upon some continuous space on top of the data that faithfully encodes the topology and the geometry of the underlying shape; one of these techniques is the simplicial complex, \cite{boissonnat:hal-01615863}. To be able to work with simplicial complexes, each element in $f^{RFM}_i$ must be transformed into a point cloud through a time delay embedding (TDE) $Z_j$.

To obtain a TDE $Z_j = \{z_j, z_{j+1}, \ldots, z_{j+w+1}\}$ of size $w$, for $j \in {1, \ldots, N-d-w+2}$ with $w$ being sufficiently large to fulfill $d \ll w \ll N$ and $d$ the number of $D=1$-dimensional objects or loops to be detected, it is necessary to generate first a cloud of $w$ points in $\mathbb{R}^d$, which changes slowly in time.

The size of $w$ is usually chosen based on the application and data. In this work, it is 30. The cloud of points consist of the vectors ${z_1, z_2, \ldots, z_{N-d+1}}$ embedded in $\mathbb{R}^d$. They are the result of a transformation of a time series $X = \{x_1, x_2, \ldots, x_N\}$ for the periods $i = 1, 2, \ldots, N$ into a sequence of $(N-d+1)$ $d$-dimensional delay coordinate vectors $z_{N-d_1}=(x_{N-d+1}, x_{N-d+2}, \ldots, x_{N})$. 
\subsection{Rips Complex}\label{subsec:point_cloud}
Once a point cloud has been obtained, Rips filtration, a popular algorithm in TDA, is used. The objective is to obtain death and birth complexes. 
The point-cloud $Z_j \subseteq \mathbb{R}^d$ for $j \in \{1, \ldots, N-d-w+2\}$ can be associated to a topological space through the so-called Vietoris-Rips simplicial complex, $R(Z, \epsilon)$, \cite{Gidea2017-em}.
This is a form of filtration $R(Z, \epsilon) \subseteq R(Z, \epsilon')$ whenever $\epsilon \leq \epsilon'$, if two conditions are met.

The first condition is that for each $j\in\{1,2,\ldots,k\}$, a $k$-simplex of vertices $\{z_{i_1}, \ldots, z_{i_k}\}$ is part of $R(Z, \epsilon)$.
The second one requires that $d(z_{i_j}, z_{i_l}) \leq \epsilon$, the mutual distance between any pair of its vertices, is less than $\epsilon $ for all $z_{i_j}, z_{i_l} \in \{z_{i_1}, \ldots, z_{i_k}\}$.

\subsection{Homology}\label{subsec:homology}
The k-dimensional homology $H_k(R(Z,\epsilon))$ of the Vietoris-Rips simplicial complex with coefficients in some field must be computed. 
This research works with the 1-dimensional homology $\{H_1(R(Z,\epsilon))\}_\epsilon))$ corresponding to independent loops in $R(Z,\epsilon)$. 

The filtration property of the Rips complexes has as a consequence that for each $k$, $H_k(R(Z,\epsilon)) \subseteq H_k(R(Z,\epsilon'))$ whenever $\epsilon \leq \epsilon'$. 

\subsection{Persistence Diagram}\label{subsec:persistence_diagram}
These inclusions determine canonical homomorphisms, a family of induced mappings. As a result, there exists a pair of values $\epsilon_1 < \epsilon_2$ for each non-zero $k$-dimensional homology class. This allows to obtain the pair $(b_\alpha, d\alpha)$ representing the 'death' and 'birth' indices of $\alpha \in H_k(R(Z,\epsilon_1))$. 
The class $\alpha$ is 'born' at $b_\alpha := \epsilon_1$ and 'dies' at $d_\alpha = \epsilon_2$.

This information can be encoded in a persistence diagram $P_k$ consisting of a point $p_\alpha = p_\alpha(b_\alpha, d_\alpha) \in \mathbb{R}^2$ together with its multiplicity $\mu_\alpha = \mu_\alpha(b_\alpha, d_\alpha)$. 
This equals to the number of classes $\alpha$ that are born at $b_\alpha$ and die at $d_\alpha$. The multiplicity is finite, since the simplicial complex is also finite. 

In addition, $P_k$ contains all points in the positive diagonal of $\mathbb{R}^2$, they represent all trivial homology generators that are born and instantly die at every level. Each point in the diagonal has a infinite multiplicity.

\subsection{Persistence Landscape}
The space of persistence diagrams can be embedded into a Banach space, for example using a persistence landscape. This consists of sequences of functions in the Banach space $L^p(\mathbb{N}\times\mathbb{R})$. 

For each $(b_\alpha, d_\alpha) \in P_k$, one must define a piece-wise linear function for the intervals $ (b_\alpha, \frac{b_\alpha+d_\alpha}{2}]$, $( \frac{b_\alpha+d_\alpha}{2}, d_\alpha)$ and $(b_\alpha, d_\alpha)$. 
A sequence of functions $\lambda = (\lambda_k)_{k\in \mathbb{N}}$, where $\lambda_k:\mathbb{R} \rightarrow [0;1]$ is associated to the persistence diagram $P_k$.

These sequence of functions are defined as
\begin{align*}
\lambda_k(x)=k\text{-max}{f_{(b_\alpha, d_\alpha)}(x) | (b_\alpha, d_\alpha) \in P_k}
\end{align*}

where $k$-max is the $k$-t largest value of a function.

\subsection{Norm of $\lambda$}
\begin{figure}[!ht]
  \begin{center}
    \includegraphics[width=\columnwidth]{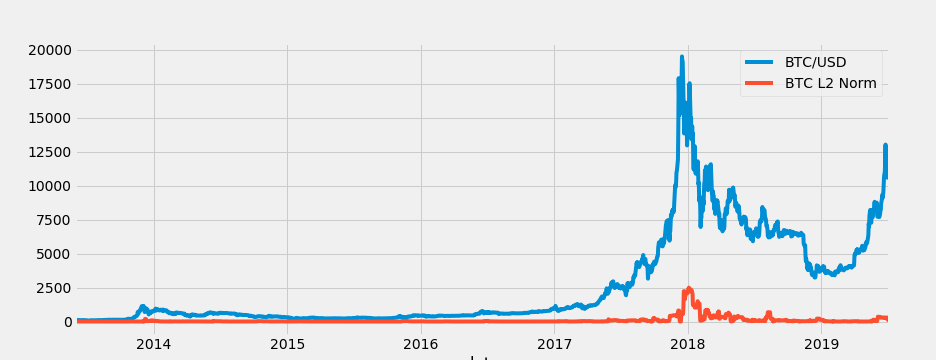}
  \end{center}
  \caption{L2-Norm of persistence landscape of Bitcoin vs USD value of Bitcoin for the period 2013-05-01 to 2019-06-30}
  \label{fig:tda:btc_l2_norm_20190501_20190701}
\end{figure}

The persistence landscapes form a subset of the Banach space $L^p(\mathbb{N}\times\mathbb{R})$, where the $L^p$-norm of $\lambda$ is given by
\begin{equation*}
\| \lambda \|_p = (\sum^\infty_{k=1} \| \lambda_k \|^p_p)^{1/p}
\end{equation*}{}

with $p \geq 1$. \cite{Dlotko2019-yr} suggests to work with the $L^2$-norm, whereas \cite{Gidea2018-ah} proposes the $L^1$-norm. In this work, the $L^2$-norm is used. Various experiments showed that it better models the behavior of the cryptocurrency markets, as it is specially robust under small perturbations. This can be seen visually in \autoref{fig:tda:btc_l2_norm_20190501_20190701}, where the $\| \lambda \|_2$-norms are displayed against the USD value of Bitcoin.

\subsection{Computation of Features}\label{subsec:barcode_features}
\begin{figure}[!ht]
  \begin{center}
    \includegraphics[width=\columnwidth]{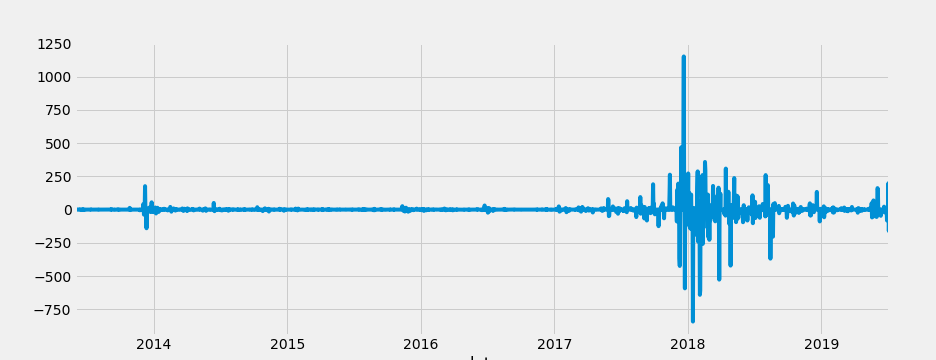}
  \end{center}
  \caption{Differences of the L2-Norms of the persistence landscape of Bitcoin for the period 2013-05-01 to 2019-06-30}
  \label{fig:tda:btc_l2_differences_20190501_20190701}
\end{figure}

The features used to calculate the portfolio allocation are based on the daily difference between two consecutive $\| \lambda \|_2$-norms, which can be defined as $\text{Diff-}L^2_{t} =\| \lambda \|_2^{t} - \| \lambda \|_2^{t-1}$ with $t \in \mathbb{Z}$, a time index. They can be visualized in \autoref{fig:tda:btc_l2_differences_20190501_20190701}, for six years of data of Bitcoin.

Similarly, this work adapts the questions presented in \autoref{section:literaturereview} to the context of cryptocurrencies with an $L^2$-norm and as such it defines them as following:
\begin{enumerate}
\item \textit{Recency}: Number of days since $\text{Diff-}L^2_{t}$ was positive.
\item \textit{Frequency}: Number of times that $\text{Diff-}L^2_{t}$ has been positive in a given period, $\sum^t_i |\text{Diff-}L^2_{i}| | \text{Diff-}L^2_{i} \geq 0$.
\item \textit{Monetary}: The cumulative value of the daily difference between two consecutive $L^2$-norms for a given time period, $\sum^t_i \text{Diff-}L^2_{i}$.
\end{enumerate}

The result is a set of three numeric features ${\textit{Recency}, \textit{Frequency}, \textit{Monetary}} \in \mathbb{N}$ or $RFM$ associated to each cryptocurrency $f_i$.

Once all features have been computed, they are combined in a single set $C$ containing all cryptocurrencies $f_i$, as such, a subscript $c in \mathbb{Z}$ is added to $RFM_c$ to indicate the features computed for each cryptocurrency $f_i$ using their respective $\text{Diff-}L^2_{i}$. Thus, $C = \{\{R_c\},\{F_c\},\{M_c\}\}$. The values of each subset of $C$ are normalized as following

\begin{equation*}
    z_i=\frac{x_i-\min(x)}{\max(x)-\min(x)}
\end{equation*} 
where $x=(x_1,...,x_n)$ and $z_i$ is the $i^{th}$ normalized element in a subset of $C$.

\subsection{TDA Portfolio}\label{section:tda_portfolio}
With the values obtained in $C$, it is possible to create a portfolio allocation based on the ranking of each cryptocurrency generated by the normalization of the values in $C$. The respective allocation is done by summing the normalized values $R^z_c + F^z_c + M^z_c$ with $z$ being a subscript to indicate that the value was normalized and $c$ identifying the respective cryptocurrency and normalizing the result. This can be called $\text{score}_c = \text{sum}(C^z_c)$. The allocation is done by dividing this score by the total number of cryptocurrencies whose $\text{score}_c \geq 0$. Thus $\text{allocation}_c = \frac{{score}_c}{|C^z_c|}$ where $C^z_c \geq 0 $.

\begin{figure}[!ht]
  \begin{center}
    \includegraphics[width=\columnwidth]{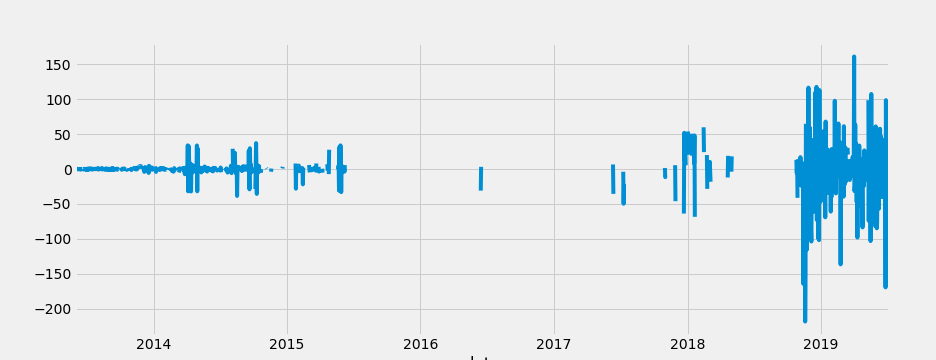}
  \end{center}
  \caption{Log returns of 1561 cryptocurrencies for the period 2013-05-01 to 2019-06-30}
  \label{fig:tda:market_log_returns}
\end{figure}

\begin{figure}[!ht]
  \begin{center}
    \includegraphics[width=\columnwidth]{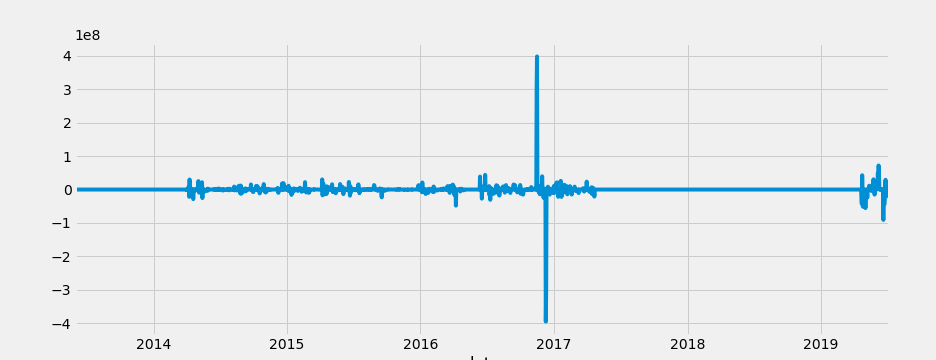}
  \end{center}
  \caption{Differences of the L2-Norms of the persistence landscape of 1561 cryptocurrencies for the period 2013-05-01 to 2019-06-30}
  \label{fig:tda:market_l2_differences}
\end{figure}

\begin{figure}[!ht]
  \begin{center}
    \includegraphics[width=\columnwidth]{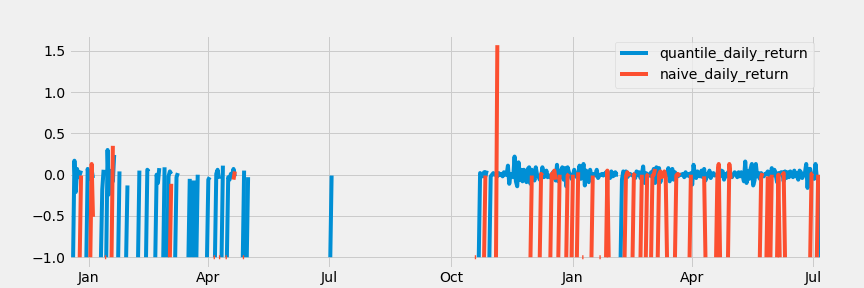}
  \end{center}
  \caption{Portfolio returns between 'TDA Portfolio' and 'Naive Portfolio' for the period 2013-05-01 to 2019-06-30}
  \label{fig:tda:btc_log_return}
\end{figure}

\section{Methods}
\label{sec:methods}
In addition to \textit{TDA for Portfolio Management of Cryptocurrencies} described in \autoref{section:tda_portfolio}, a traditional method for portfolio allocation is used as a benchmark. It allocates the investment amount equally over all cryptocurrencies, $1/N$ with $N$ being the total number of cryptocurrencies available. This method is denoted as 'Naive Allocation'

\section{Discussion \& Learnings}
The results between the 'TDA Portfolio and the 'Naive Portfolio' can be seen for the period 2017-12-17 to 2019-07-05 in \autoref{fig:tda:btc_log_return}. The start date was selected as this was a period of intense interest among general audiences for cryptocurrencies. This was very close to the peak of \$19,500 per Bitcoin. As such, this encouraged many new investors to enter the market. Since then, results have diverged significantly and this can be seen in \autoref{fig:tda:btc_log_return}. 

The returns are for both portfolios negative. However, for the 'Naive Portfolio' the performance was significantly worse. This can be seen more in detail in \autoref{tab:portfolio_returns}, where the naive porfolio results in significant monthly loses, whereas the TDA portfolio offers a better performance.

\begin{table}[H]
\caption{Adjusted Performance of TDA Porfolio vs Naive Porfolio over the period January 2018 to July 2019. Lower number denotes a worse performance of the portfolio}
\centering
\label{tab:portfolio_returns}
\begin{tabulary}{\linewidth}{CCC}
    \toprule
Date & TDA Portfolio & Naive Portfolio \\ \hline
2017-12&-2.88&-4.03 \\ 
2018-01&-6.01&-11.03 \\
2018-02&-7.78&-5.93 \\
2018-03&-10.10&-10.11 \\
2018-04&-7.94&-12.02 \\ 
2018-05&-5.03&-9.0 \\
2018-06&-2.0&-6.0 \\ 
2018-07&-2.01&-9.0 \\ 
2018-08&0.0&-7.0 \\
2018-09&0.0&-7.0 \\
2018-10&-1.96&-9.01 \\
2018-11&0.06&-7.43 \\
2018-12&0.08&-9.0 \\
2019-01&0.05&-10.93 \\
2019-02&-0.99&-7.95 \\
2019-03&0.05&-8.93 \\
2019-04&0.01&-11.81 \\
2019-05&0.12&-10.03 \\
2019-06&0.09&-8.97 \\
2019-07&-1.54&-1.98 \\
 \bottomrule
 \end{tabulary}
\end{table}

Another benefit of using a TDA-based approach can be seen in \autoref{fig:tda:market_l2_differences} and \autoref{fig:tda:market_l2_differences}. The daily returns for the market are depicted both in the form of the norms as well as the log returns. The log returns contain significant noise, whereas the TDA norms show the peak in the market in late 2017 and its later collapse.

\subsection{On deploying in production}
Due to the scope of this study, the proposed methods have not yet been deployed in production. From a technical perspective, this study was done on a server with 256GB RAM, 40 CPUs, 2 threads per core, 20 cores per socket and one socket.

\section{Takeaways for the practitioner}

\cite{Rivera-Castro2019-qk} already showed that TDA can be used successfully for commercial time series and \cite{Gidea2018-ah} showed that it can also be used to analyze cryptocurrencies. This work takes these ideas and combines them together with a popular managerial framework to provide an alternative for asset allocation using exclusively time series data collected from public sources.

A set of experiments were carried to assess the suitability of this system. The results show that using the $L^2$-norms provides more insights and is more effective than using a traditional portfolio method.

However, time series with TDA is not yet suited for large data sets. Computing the homologies is expensive. Another limitation of this method is the lack of established best practices for defining the size of the sliding window to generate the point cloud. This can be especially difficult whenever dealing with sparse time series data. Thus, in order to improve the method, two bottle necks need to be addressed. On one side, the computation using TDA methods. On the other side, develop methods to optimize the window size and related parameters. 

This study provides an implementation for all experiments to be used out-of-the-box. Finally, the data set made available is unique and novel. There are no other public data sets for cryptocurrencies that are comparable.

\section{Conclusion}
This work presents an approach for portfolio management of cryptocurrencies using TDA and time series data. Given the unique characteristics of cryptocurrencies, traditional tools for portfolio management are not well-suited.

As a further line of work, this work will seek to cover very large data sets with intra-day data. An additional line is to adapt other popular frameworks for portfolio allocation to the context of TDA methods. 

Overall, TDA is a nascent field specially in combination with analysis of cryptocurrencies. As the field grows in popularity and new business applications appear, it is to be expected that TDA will become an essential tool for the practitioner.

\section{Reproducibility}
\subsection{Topological Time Series Clustering}\label{repro:topo_rfm}
The 'TDA Portfolio' pipeline consists of the following steps:
\begin{enumerate}
\item As a first step, the time series must be converted using sliding windows. The objective is to generate delay embeddings that can be projected as a point cloud.
\item Once the point clouds have been obtained, Rips filtration, a popular algorithm in TDA, is used, with the objective of obtaining death and birth complexes. These processes can be visualized in the form of persistence diagrams. The points of interest are those outside of the diagonal.
\item As a fourth step, barcode diagrams are generated for both 0- and 1- dimensional homologies. The focus is on the 1-dimensional homologies (loops). The 0-dimensional ones do not provide relevant information.
\item This information is used to generate the persistence landscapes.
\item One computes the $L^2$-norm of each persistence landscape, and as a result, one obtains a time series of norms.
\item The difference of the $L^2$-norms is computed for each cryptocurrency.
\item Features around the concept of RFM are computed for each cryptocurrency
\item The features are normalized and added to obtain a final score
\item The score is used to define the amount to be allocated to each asset.
\item Portfolio returns are calculated for the period 2017-12-17 to 2019-07-05
\end{enumerate}

\bibliographystyle{IEEEtran}
\bibliography{bib}

\begin{thebibliography}{10}
\providecommand{\url}[1]{#1}
\csname url@samestyle\endcsname
\providecommand{\newblock}{\relax}
\providecommand{\bibinfo}[2]{#2}
\providecommand{\BIBentrySTDinterwordspacing}{\spaceskip=0pt\relax}
\providecommand{\BIBentryALTinterwordstretchfactor}{4}
\providecommand{\BIBentryALTinterwordspacing}{\spaceskip=\fontdimen2\font plus
\BIBentryALTinterwordstretchfactor\fontdimen3\font minus
  \fontdimen4\font\relax}
\providecommand{\BIBforeignlanguage}[2]{{%
\expandafter\ifx\csname l@#1\endcsname\relax
\typeout{** WARNING: IEEEtran.bst: No hyphenation pattern has been}%
\typeout{** loaded for the language `#1'. Using the pattern for}%
\typeout{** the default language instead.}%
\else
\language=\csname l@#1\endcsname
\fi
#2}}
\providecommand{\BIBdecl}{\relax}
\BIBdecl

\bibitem{Corbet_undated-au}
S.~Corbet, B.~M. Lucey, A.~Urquhart, and L.~Yarovaya, ``Cryptocurrencies as a
  financial asset: A systematic analysis.''

\bibitem{Corbet2018-td}
S.~Corbet, A.~Meegan, C.~Larkin, B.~Lucey, and L.~Yarovaya, ``Exploring the
  dynamic relationships between cryptocurrencies and other financial assets,''
  \emph{Econ. Lett.}, vol. 165, pp. 28--34, Apr. 2018.

\bibitem{Platanakis2019-jk}
E.~Platanakis and A.~Urquhart, ``Portfolio management with cryptocurrencies:
  The role of estimation risk,'' \emph{Econ. Lett.}, vol. 177, pp. 76--80, Apr.
  2019.

\bibitem{Zhang2015-wa}
Y.~Zhang, E.~T. Bradlow, and D.~S. Small, ``Predicting customer value using
  clumpiness: From {RFM} to {RFMC},'' \emph{Marketing Science}, vol.~34, no.~2,
  pp. 195--208, Mar. 2015.

\bibitem{Gidea2018-ah}
M.~Gidea, Y.~A. Katz, P.~Roldan, D.~Goldsmith, and Y.~Shmalo, ``Topological
  recognition of critical transitions in time series of cryptocurrencies,''
  Jun. 2018.

\bibitem{Akcora_undated-ew}
C.~G. Akcora, A.~K. Dey, Y.~R. Gel, and M.~Kantarcioglu, ``{PAKDD} :
  Forecasting bitcoin price with graph chainlets.''

\bibitem{Akcora2019-rw}
C.~G. Akcora, Y.~Li, Y.~R. Gel, and M.~Kantarcioglu, ``{BitcoinHeist}:
  Topological data analysis for ransomware detection on the bitcoin
  blockchain,'' Jun. 2019.

\bibitem{Akcora2018-gl}
C.~G. Akcora, M.~Kantarcioglu, and Y.~R. Gel, ``Blockchain data analytics,''
  2018.

\bibitem{books/daglib/0025666}
H.~Edelsbrunner and J.~Harer, \emph{Computational Topology - an
  Introduction.}\hskip 1em plus 0.5em minus 0.4em\relax American Mathematical
  Society, 2010.

\bibitem{boissonnat:hal-01615863}
\BIBentryALTinterwordspacing
J.-D. Boissonnat, F.~Chazal, and M.~Yvinec, \emph{{Geometric and Topological
  Inference}}.\hskip 1em plus 0.5em minus 0.4em\relax {Cambridge University
  Press}, 2018, cambridge Texts in Applied Mathematics. [Online]. Available:
  \url{https://hal.inria.fr/hal-01615863}
\BIBentrySTDinterwordspacing

\bibitem{chazal2017introduction}
F.~Chazal and B.~Michel, ``An introduction to topological data analysis:
  fundamental and practical aspects for data scientists,'' 2017.

\bibitem{doi:10.1177/1094670506293810}
S.~Gupta, D.~Hanssens, B.~Hardie, W.~Kahn, V.~Kumar, N.~Lin, N.~Ravishanker,
  and S.~Sriram, ``Modeling customer lifetime value,'' \emph{Journal of Service
  Research}, vol.~9, no.~2, pp. 139--155, 2006.

\bibitem{Zaki2016-oj}
M.~Zaki, D.~Kandeil, A.~Neely, and J.~R. McColl-Kennedy, ``The fallacy of the
  net promoter score: Customer loyalty predictive model,'' 2016.

\bibitem{Wubben2008-yf}
M.~W{\"u}bben, \emph{\BIBforeignlanguage{en}{Analytical {CRM}: Developing and
  Maintaining Profitable Customer Relationships in {Non-Contractual}
  Settings}}.\hskip 1em plus 0.5em minus 0.4em\relax Gabler Verlag, Oct. 2008.

\bibitem{Rivera-Castro2019-qk}
R.~Rivera-Castro, P.~Pilyugina, A.~Pletnev, I.~Maksimov, W.~Wyz, and
  E.~Burnaev, ``Topological data analysis of time series data for {B2B}
  customer relationship management,'' 2019, in press.

\bibitem{Gidea2017-em}
M.~Gidea and Y.~Katz, ``Topological data analysis of financial time series:
  Landscapes of crashes,'' Mar. 2017.

\bibitem{Dlotko2019-yr}
P.~D{\l}otko, W.~Qiu, and S.~Rudkin, ``Cyclicality, periodicity and the
  topology of time series,'' May 2019.

\end{thebibliography}

\end{document}